\documentclass[12pt]{article}
\usepackage{graphicx,epsfig}
\def\ni{\noindent}
\begin{document}

\begin{center}
{\Large
Simulations of viscous shape relaxation in shuffled foam clusters}

\vskip .5cm
Gilberto L. Thomas$^1$, Jos\'e C.M. Mombach$^2$, Marco A. P.
Idiart$^1$, \\ C. Quilliet$^3$,  F. Graner$^3$

\vskip .3cm

{\footnotesize
$^1$ Instituto de F\'{\i}sica, Universidade Federal do Rio Grande do
Sul,  \\ C.P. 15051, 91501-970 Porto Alegre, Brazil.

$^2$ Laborat\'orio de Bioinform\'atica e Biologia Computacional \\
Centro de Ci\^encias Exatas e Tecnol\'ogicas, Universidade do
Vale do Rio dos Sinos, \\ Av. Unisinos, 950 93022-000 S\~ao Leopoldo, 
RS, Brazil

$^3$ Laboratoire de Spectrom\'etrie Physique, BP 87, 38402 St. Martin
d'H\`eres Cedex, France.

}
\end{center}

\abstract

{\footnotesize
We simulate the shape relaxation of foam clusters and compare them
with the time exponential expected for Newtonian fluid. Using
two-dimensional Potts Model simulations, we  artificially create
holes  in a foam cluster and shuffle it by applying shear strain
cycles. We reproduce the experimentally observed time exponential relaxation
of cavity shapes in the foam as a function of the number of strain
steps. The cavity rounding up results from local rearrangement of
bubbles, due to the conjunction of both a large applied strain and
local bubble wall fluctuations.}

\vskip .5cm
\noindent Keywords: {\bf  two-dimensional foam cluster, cellular
fluid, hydrodynamic relaxation, Potts model. }

\vskip .5cm
\section{Introduction}

Liquid foams, and especially two-dimensional ones, are easy to image
and manipulate, and hence can be  a quite efficient model for other complex
systems \cite{wh99} and,
in particular, for biological cell aggregates \cite{gg93,Honda04,DR97}.

Biological cell aggregates and foams share a common point: they  are 
``cellular fluids";
they can flow while their constituent cells  rearrange and change 
neighbors through
relative local movements \cite{wh99}. When out of equilibrium, their mechanical
behavior displays elastic, plastic or viscous responses.

Interestingly, the rounding up of cell aggregates observed in biology 
is similar to
the coalescence or rounding up of liquid droplets. Gordon {\it et
al.} \cite{Gordon} made a formal analogy with the rheology of viscous
liquids, showing that tissue surface tensions drive these processes,
while tissue viscosity resists them.  Recently, Rieu and Sawada
\cite{RS02} showed that a hydrodynamic laws also apply to the
rounding-up of two-dimensional (2D) aggregates of mixed ectodermal
and endodermal
{\it hydra} cells in various proportions. They obtained
time-exponential relaxations for rounding up of aggregates.
Corresponding simulations \cite{zecarlos} have investigated the
relaxation of biological cell aggregates of elliptical shapes.

What is common between the relaxation of biological cell aggregates and foams,
and how does it  differ from that of an usual Newtonian fluid?
To address this question, the companion paper \cite{qidby04} studies
the relaxation of a foam cluster, and in the present paper we perform
the corresponding simulations using Cellular Potts Model \cite{gg93,gg92} .

Concerning the mechanism of relaxation a foam has an important
difference with biological cell aggregates.
Cell membranes fluctuate actively, with a r.m.s. movement
equivalent to that produced by a very high effective temperature
(much higher than room temperature); so that cells explore their
neighborhood and can move.
The macroscopic configuration of a foam, on the other hand, arises
from the local equilibrium of the bubbles' surface tensions and
pressures. A foam never reaches its global energy minimum since the
energy barriers between local equilibrium states, attainable by
moving passive bubbles around, are much higher than the order $k_B T$
thermal energy fluctuations at room temperature.
At bubble scales, we can consider a foam  to be at zero temperature,
so that the foam configuration falls and is trapped in an energy
minimum; we do not expect to observe rounding up of foams at room
temperature.

In order to circumvent that we introduce energy in the foam through
successive shear cycles \cite{qidby04,visnoff03}.
The external force shears the foam, rearranging bubbles, and
exploring new energy configurations, and relaxing into more stable
configurations. Our goal is to determine if the relaxation laws we
obtain by this method are the same as those for aggregates of
biological cells, \emph{i.e.}, if mechanical stimulation is
equivalent to an effective temperature.
However, experimentally, applying external mechanical strain turned
much easier for foam cluster with a hole in it (Fig. 1) than for a
cluster which does not touches the box walls. Thus we address here
the cavity relaxation, as if it was the ``negative" of the aggregate
relaxation \cite{RS02,zecarlos}.
We also study the center of mass movement of cavities of different
sizes in the foam  to compare with the experimental observations in 
\cite{qidby04}.

\section{The Model}

The 2D model
reflects the quasi-2D experimental foam of Ref. \cite{qidby04}, see Fig. 1.
   We use the extended large-$Q$ Potts model, which allows large
numbers of bubbles,   fixed bubble
areas, and large foam distortions
\protect\cite{jiangpre}.
   This model represents the spatial structure of the foam as follows.
We consider  a 2D square grid. The model treats a foam on a 2D
lattice by assigning an integer
index to each lattice site. The value at a grid site ($i, j)$ is
\textit{$\sigma $} if the site lies inside bubble \textit{$\sigma$}.
Each bubble   of the foam is thus the connected set of grid sites
with the same index \textit{$\sigma $}. Each bubble is thus labelled
by this index  \textit{$\sigma $} and occupies many grid sites.

Each
pair of neighbors having unmatched indices determines a bubble wall
and contributes to the bubble
wall surface energy. The prefactor $J$ is the surface tension (which
we can here take equal to 1 without loss of generality).
Bubble areas are the number of lattice sites for each
index; an area constraint
keeps the bubble areas constant.
The actual bubble area is $A(\sigma)$ and the target bubble area is
$A_t$, the same for each bubble. The prefactor $\lambda$ is the
compressibility of the bubble. The Hamiltonian thus writes:
\begin{equation}
\mathop H = \sum_{{(i,j),(i',j')}}J(\tau(\sigma (i,j)),\tau'(\sigma
'(i',j')))(1 - \delta _{\sigma (i,j),\sigma
'(i',j')} ) + \sum\limits_\sigma \lambda(\tau) (A(\sigma) -
A_t)^2  \,\, . \nonumber
\end{equation}
Here the second sum is over all bubbles, while the first sum is over
all pairs of neighboring sites of coordinates $(i, j)$ and $(i',
j')$. We use the 5th nearest neighbors to decrease the effect of the
anisotropy of a discrete lattice \cite{holm}.
$J(\tau(\sigma (i,j)),\tau'(\sigma '(i',j')))$ denotes the
interaction energy between the neighboring sites $(i,j)$ and $(i',
j')$.
We treat the cavity as a large single bubble; we thus introduce the index
    $\tau$ which is the bubble type: the cavity  has $\tau = 0$, while
all the bubbles are of type $\tau = 1$. Since all bubbles have a
fixed area, and the total area is fixed too, a constraint on the
cavity would be redundant (leading to numerical instabilities), so we
do not enforce it (we set  $\lambda(0)$ to 0).

Energy minimization through Metropolis
dynamics~\cite{gg93,gg92} minimizes the total wall energy (sum of
bubble perimeters).
In the Monte Carlo dynamics, at each step, we select a random grid
site ($i$,$j$). An attempt to change its index from \textit{$\sigma
$} to \textit{$\sigma $'} is tried (where \textit{$\sigma $'} is the
index of an arbitrary lattice site among one of its first neighbors)
and is performed with the probability,:
\begin{eqnarray}
P(\sigma(i,j) \to \sigma '(i,j)) = \left
   \{ \begin{array}{ll}
   \mbox{exp}(-\Delta {H}/kT) & \mbox{if }
\,\, \Delta \mbox{$H$} > 0, \, \mbox{or} \\
   1 & \mbox{if }\,\,
\Delta \mbox{$H$} \le 0,
\end{array}\right.
\end{eqnarray}
where   $\Delta {H}$ is the energy gain the change produces.
Each such move corresponds to bubble \textit{$\sigma $'} displacing
bubble \textit{$\sigma $} by one lattice site. $T>0$ is a fluctuation
temperature corresponding to the amplitude of bubble wall
fluctuations. The time unit, a Monte Carlo step (MCS), corresponds by
definition to as many attempts as the number of lattice sites.

\section{Cavity relaxation}

Simulations run on square grid containing about 20 x 20
bubbles of the same size (either 63 or 99 lattice sites), with a
centered cavity whose size we measure in number of bubbles (ratio of
cavity area to bubble area).

To   shear the foam, we proceed in the following way; which closely
mimics the experimental set-up (Fig. 1).
  From the initial state, we periodically displace the lower part of
the grid horizontally up to the desired maximum angle, and in the
opposite direction as well. We  slide each row by interpolating
between the lower (moving) and upper (fixed) sides. This
procedure distorts each bubble, which can
change its shape and its neighbors. Then we allow the foam to relax.

We observed that, at all angles we tried (up to $ 30^\circ$), at
zero temperature this procedure alone does not cause rounding.
Therefore at each position shown in Fig. 1 we relax at finite
temperature for 100 MCS, then anneal at  $T=0$ for the same amount of
time. We choose a temperature small enough so that the thermal
treatment by itself does not cause rounding. Since one shear cycle is
performed in four such sub-steps, it takes 800 MCS.

Defining the cavity eccentricity as $e=1- \mbox{minor axis/major
axis}$, Fig. 2 shows the relaxation for different maximal
displacement angles. We chose two different initial cavity shapes:
both shapes (rectangular and ellipsoidal) have the same behavior.
   Higher shear result in quicker relaxation. In all cases, the
semi-log plot is close to a straight line: the relaxation is close to
exponential in time, enabling us to define a
   characteristic time ($\tau_c$).

Fig. 3 shows   $\tau_c$ for rounding of foams with two different
bubble sizes (63 and 99 lattice sites), but with the same cavity /
bubble size ratio, and for different maximal angles. Rounding is
faster for larger bubbles and for larger maximal angles.

We have also investigated the behavior of cavities of different
sizes. Fig. 4 shows a large variation in cavity sizes. Initially the 
rounding is exponential, but it later slows down, especially for
small cavities for which it saturates, in agreement with Ref.
\cite{qidby04}. Fig. 5 shows the rounding characteristic time for this
situation. We can see that smaller cavities round faster.
\section{Cavity diffusion}

Motivated by the experimental observations \cite{qidby04}, we
performed a set of measurements to
    investigate the diffusion of the cavity as a result of both
shearing and increasing the temperature. Fig. 6 shows the cavity
center of mass position during shear cycles for different cavity
sizes. As expected, the smaller the cavity, the more it moves.

To test whether this movement corresponds to a diffusion,
Fig. 7 displays the quadratic center of mass displacement  for the
same cavity sizes. The answer is no: we do not observe  any
significant  diffusion. The same holds if we increase the deformation
angle (Fig. 8).

\section{Summary}

Cavities of different shapes round similarly. The combination of
deformation and finite temperature annealing in our simulations
reproduces the exponential relaxation behavior of rounding up of biological
cell aggregates. Our results have also shown that rounding is faster
for larger maximal sheared angles, for cavities in foams with larger
bubbles and for cavities with smaller sizes. Neither temperature, in
the range we choose, nor shearing alone causes rounding.
The finite temperature annealing in the simulation
might represent some other form of energy transfer, present in the
experiment, for example vibration of the foam during mechanical
shearing.  No evidence suggests that shearing the system induces
diffusion.

The next step in our investigation is to study, by experiments and
simulations, single bubble movement during shearing. If periodic
shearing represents an input of energy equivalent to a thermal bath
of finite temperature, individual bubbles could diffuse, even if the
cavity does not.

\section*{Acknowledgements}

Capes--Cofecub exchange project No. 414/03 supported this work. J. C.
M. Mombach acknowledges the partial support of HP Brazil R\&D.

\newpage
\section*{Figures}

\vskip 1cm
\begin{center}
\includegraphics[width=13cm,height=6cm]{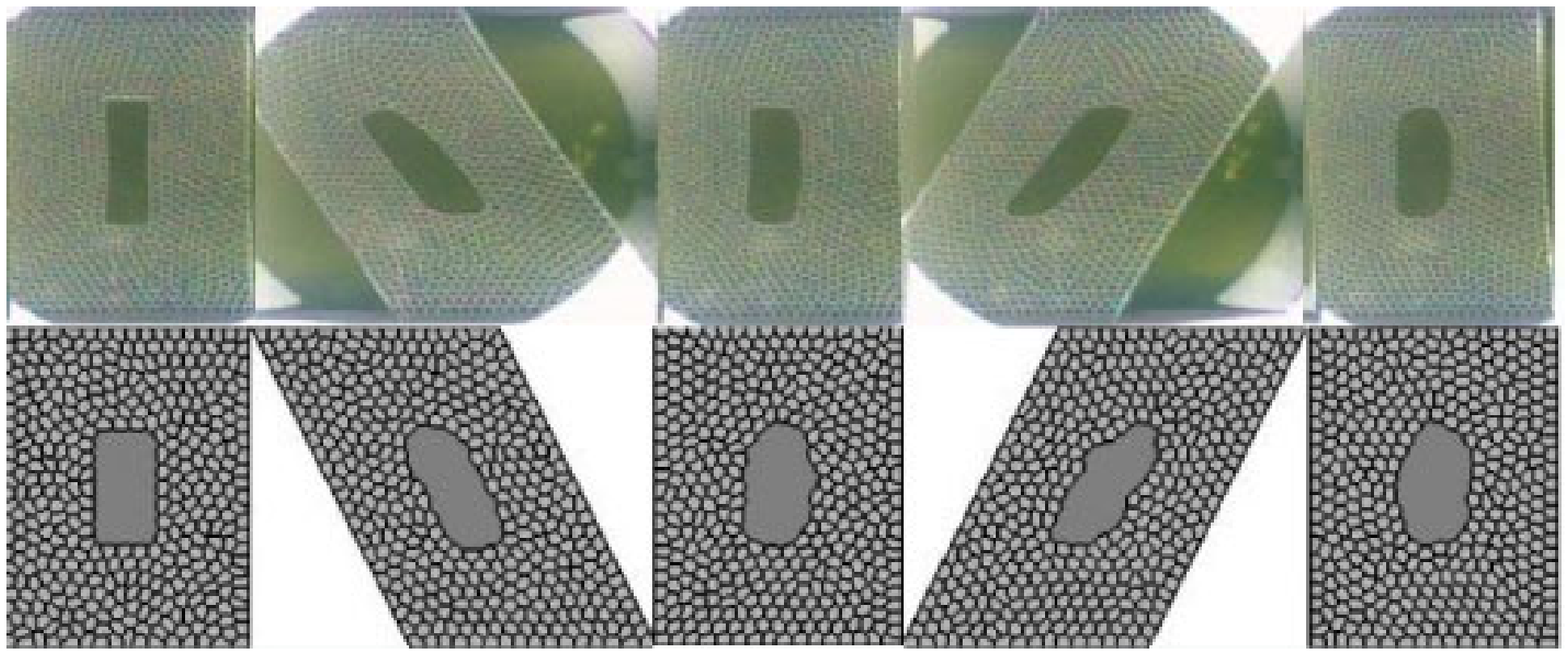}

{\bf Fig. 1}
\end{center}

\vskip 1cm
\begin{center}
\includegraphics{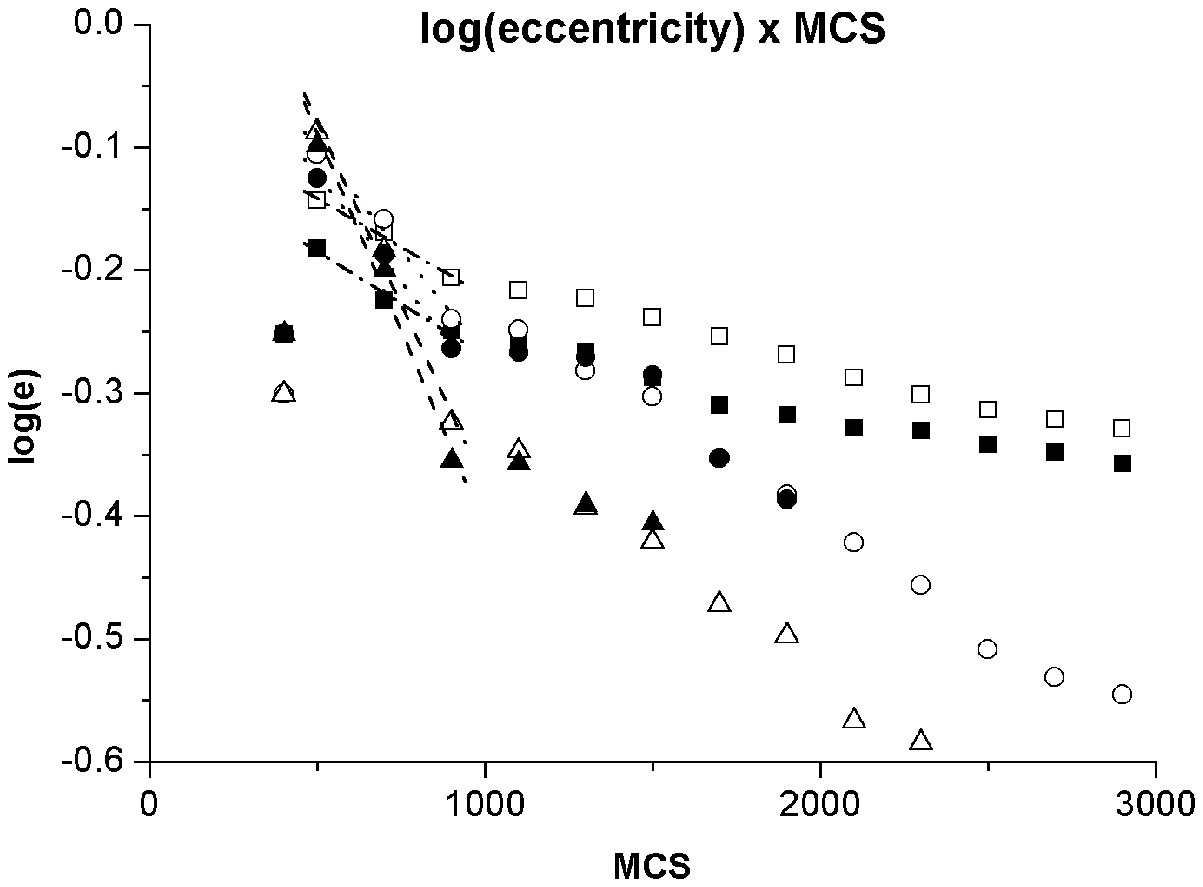}

{\bf Fig. 2}
\end{center}

\vskip 1cm
\begin{center}
\includegraphics{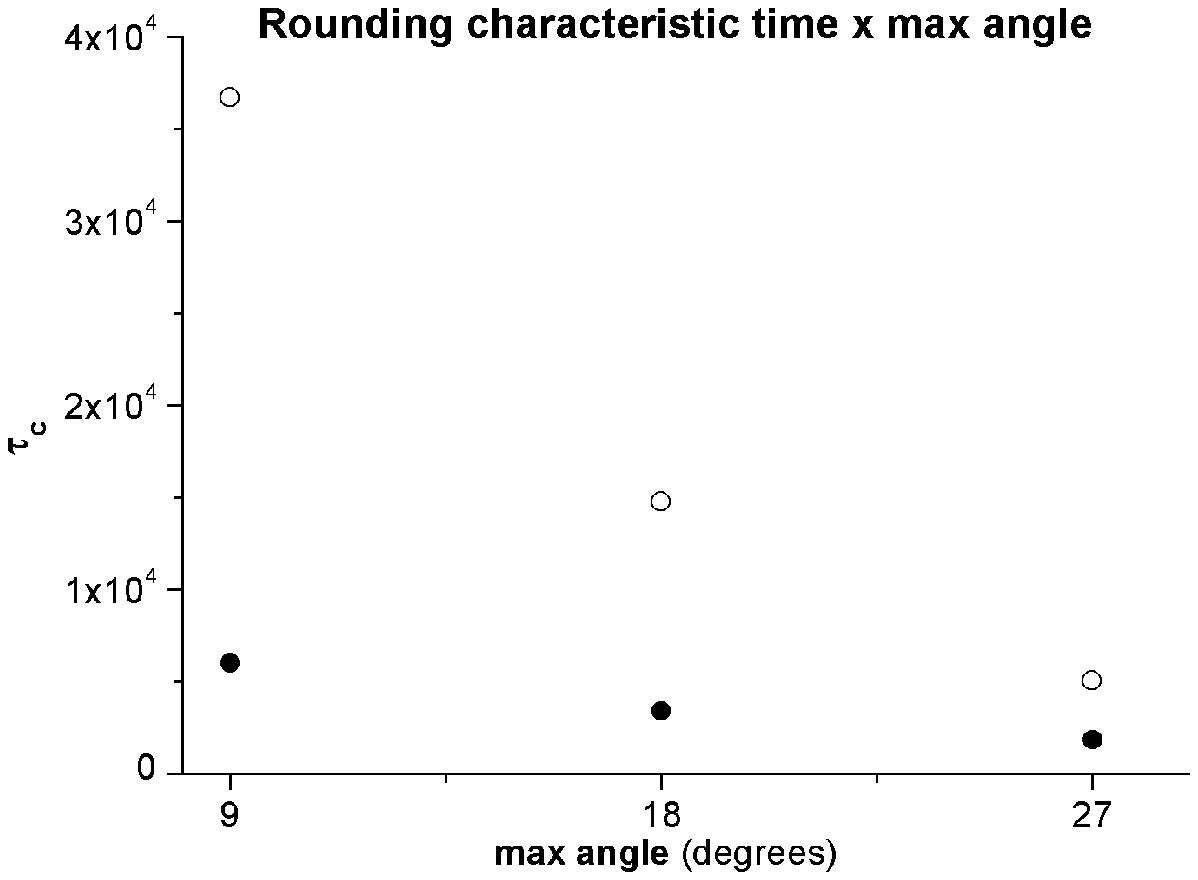}

{\bf Fig. 3}
\end{center}

\vskip 1cm
\begin{center}
\includegraphics{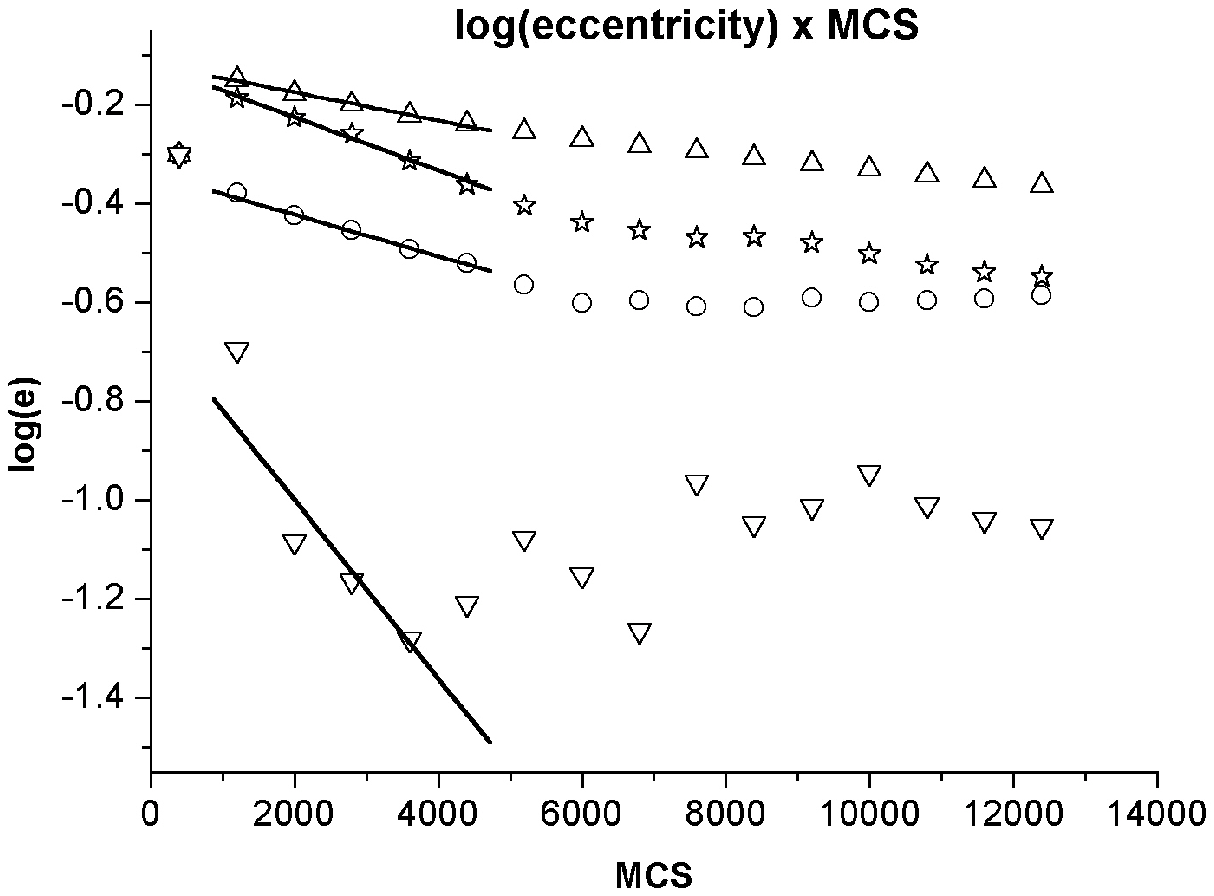}

{\bf Fig. 4}
\end{center}

\vskip 1cm
\begin{center}
\includegraphics{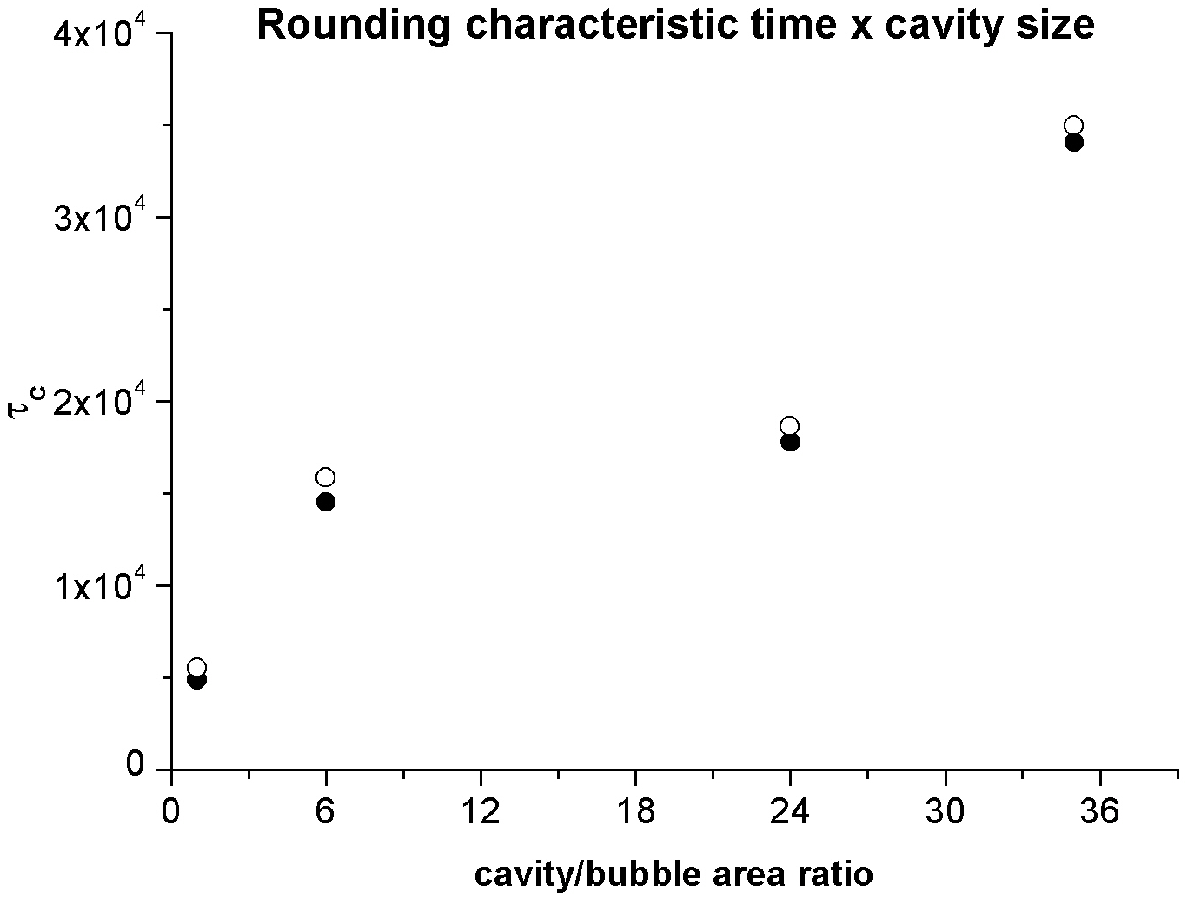}

{\bf Fig. 5}
\end{center}

\vskip 1cm
\begin{center}
\includegraphics{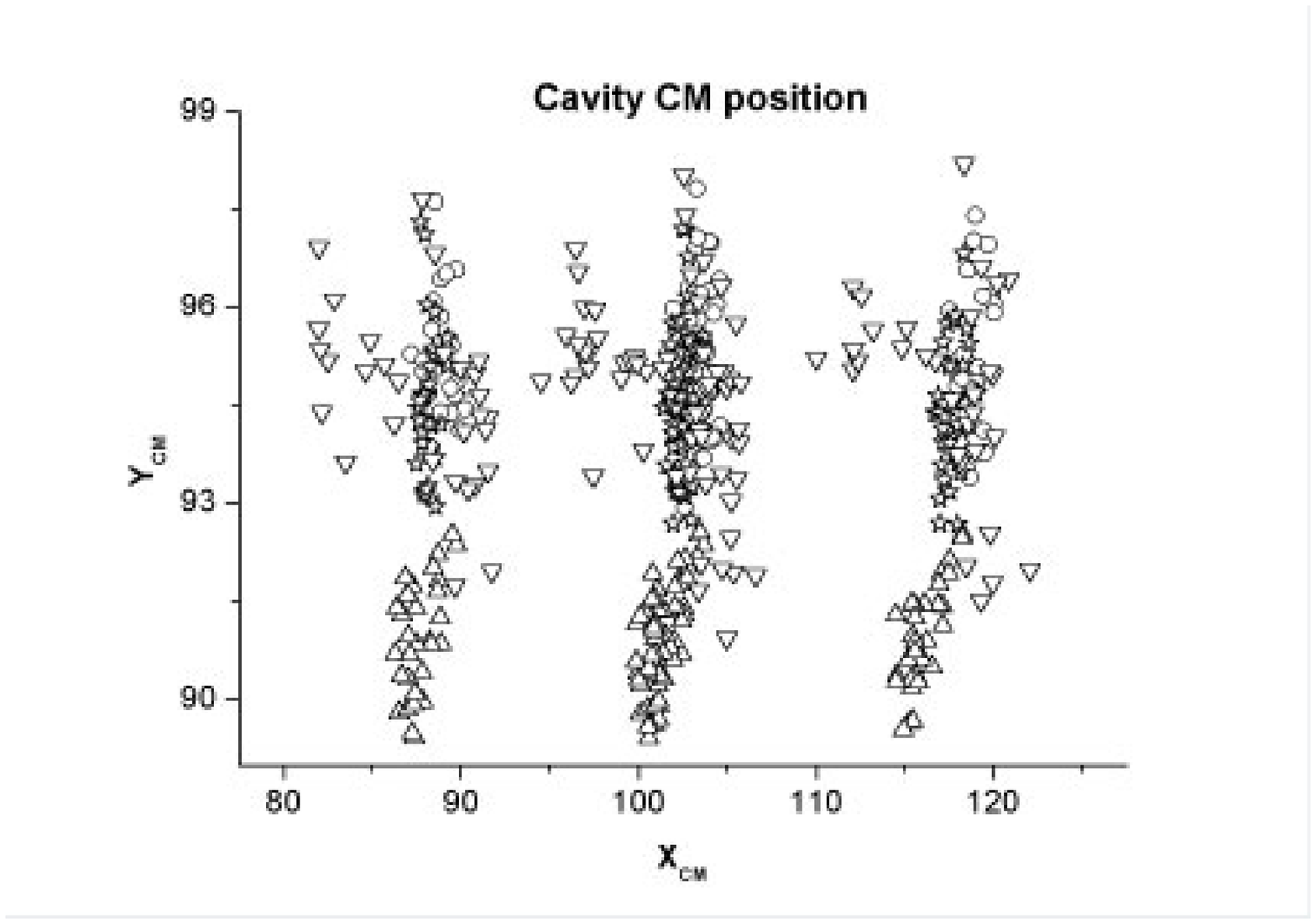}

{\bf Fig. 6}
\end{center}

\vskip 1cm
\begin{center}
\includegraphics{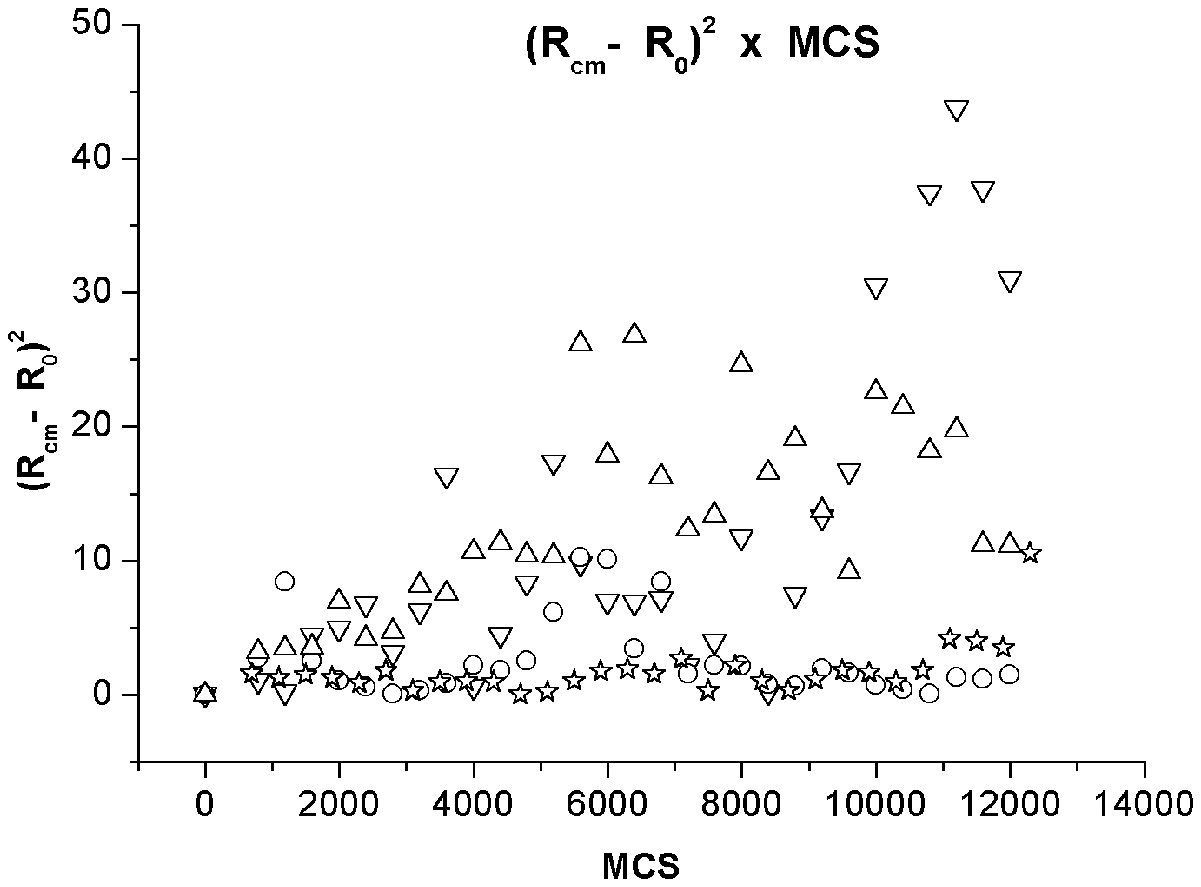}

{\bf Fig. 7}
\end{center}

\vskip 1cm
\begin{center}
\includegraphics{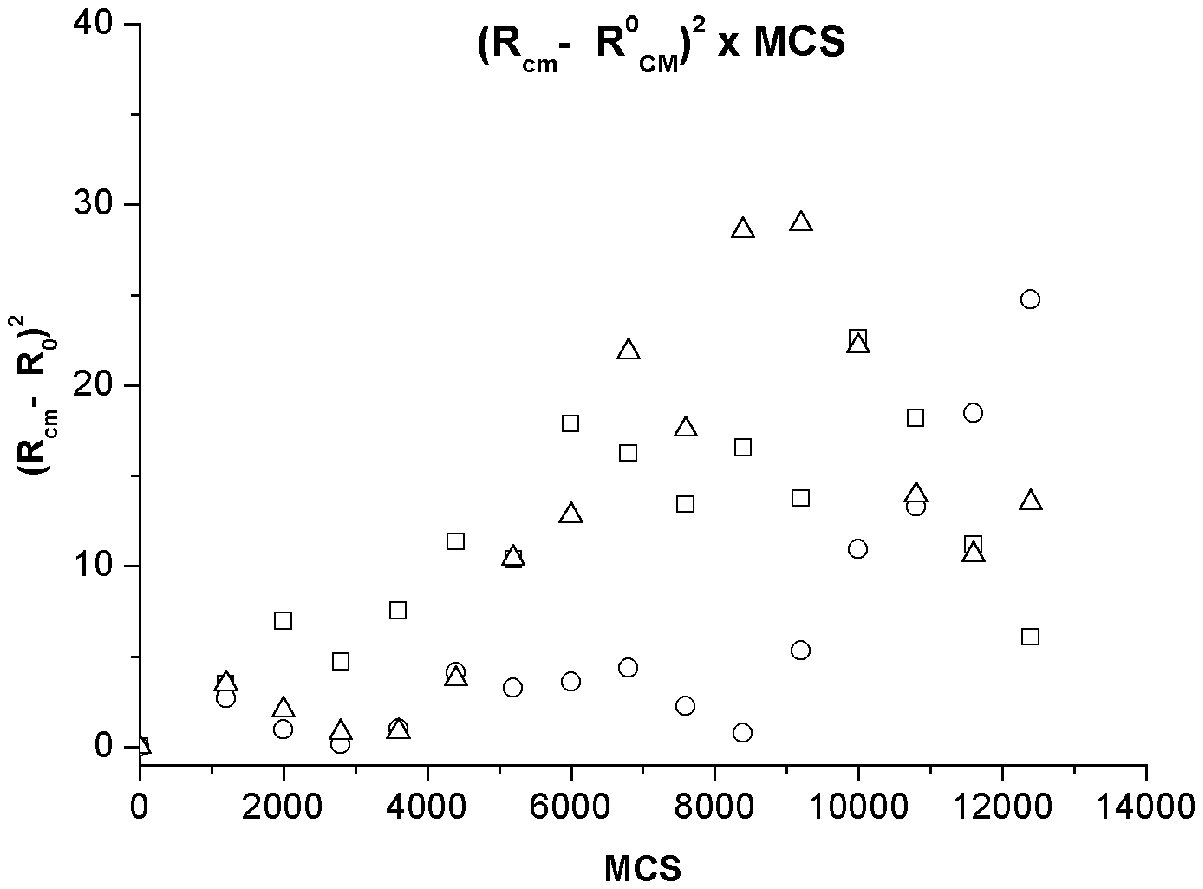}

{\bf Fig. 8}
\end{center}

\newpage
\section*{Figure captions}

\vskip 1cm
\ni{\bf Fig. 1} One shear cycle, here with initially
rectangular cavity. Top: experimental setup \protect\cite{qidby04}.
The foam is sheared back and forth, to the maximal angle, in both
directions. Bottom: corresponding  simulation (bubble size = 63
lattice sites).

\vskip 1cm
\ni{\bf Fig. 2} Time relaxation of cavity
shape: $\log(e)$ {\it versus} time (MCS) for different maximal
angles. After a transient, both rectangular and ellipsoidal cavities
round similarly. Shear is labelled by the maximal deformation angle:
9, 18 or 27$^\circ$.

\vskip 1cm
\ni{\bf Fig. 3} Characteristic relaxation
time $\tau_c$ {\it versus} maximal shear angle for two different
bubble sizes: 63 lattice sites (open circles) and 99 lattice sites (closed
circles).

\vskip 1cm
\ni{\bf Fig. 4} Cavity relaxation: $\log(e)$
{\it versus}  time (MCS) for very different cavity area / bubble area
ratios: 1 (inverted triangle), 6 (circle), 24 (star) and 35 
(triangle). Bubble size = 63, elliptical initial cavity,
deformation angle 9$^\circ$.

\vskip 1cm
\ni{\bf Fig. 5}  Characteristic relaxation
time $\tau_c$ {\it versus} cavity size. Bubble area is fixed to 63;
elliptical initial shape; deformation angle 9$^\circ$.
Note that if we consider all data taken at half cycles (closed
circles) we obtain the same results as if we perform measurements
only at integer cycles (open circles).

\vskip 1cm
\ni{\bf Fig. 6}  Cavity center of mass
position for different cavity area / bubble area ratios: 1 (inverted 
triangle), 6 (circle), 24 (star) and 35 (triangle). Bubble size = 63, 
elliptical initial cavity, deformation angle 9$^\circ$. Note the 
difference in horizontal and vertical scales.

\vskip 1cm
\ni{\bf Fig. 7}  Center of mass displacement
$(R_{CM} - R_o)^2$ {\it versus}  time (MCS) for different cavity area
/ bubble area ratios: 1 (inverted triangle), 6 (circle), 24 (star) 
and 35 (triangle).  Bubble size = 63, elliptical
initial cavity, deformation angle 9$^\circ$.

\vskip 1cm
\ni{\bf Fig.8} Center of mass displacement
$(R_{CM} - R_o)^2$ {\it versus}  time (MCS) for different maximal
sheared angle: 9 (square), 18 (circle) and 27$^\circ$(triangle). 
Cavity area / bubble area ratio = 35, bubble size = 63, elliptical 
initial cavity.

\end{document}